\documentclass[epj]{svjour}
\usepackage{graphics}
\usepackage{epsfig,latexsym}
\usepackage{dcolumn}% Align table columns on decimal point

\newcommand{\pT} {\mbox{$p_T$}}
\newcommand{\pp} {\mbox{$p$$+$$p$}}
\newcommand{\pA} {\mbox{$p$$+$$A$}}
\newcommand{\NN} {\mbox{$A$$+$$A$}}

\newcommand{\xBJ} {\mbox{x$_{BJ}$}}

\newcommand{\qbar} {\mbox{$\overline{q}$}}

\begin{document}

\title{The Future of Hard and Electromagnetic Probes at RHIC}
\author{John W. Harris}
\institute{{Physics Department, Yale University, \\
P.O. Box 208124, 272 Whitney Avenue, New Haven CT, U.S.A.
06520-8124}} %\texttt{John.Harris@Yale.edu}}
\date{Received: date / Revised version: date}
\abstract{Potential near- and long-term physics opportunities with
jets, heavy flavors and electromagnetic probes at RHIC are
presented. Much new physics remains to be unveiled using these
probes, due to their sensitivity to the initial high density stage
of RHIC collisions, when quark-gluon plasma (QGP) formation is
expected. Additional physics will include addressing
deconfinement, chiral symmetry restoration, properties of the
strongly-coupled QGP and a possible weakly-interacting QGP, color
glass condensate in the initial state, and hadronization. To fully
realize the physics prospects of the RHIC energy regime, new
detector components must be added to existing experiments, the
RHIC machine luminosity upgraded, and a possible new detector with
significantly extended coverage and capabilities added.
 % end of PACS codes
} %end of abstract
\maketitle

\section{Physics Goals at RHIC}

The primary physics goal of the Relativistic Heavy Ion Collider
(RHIC) has been to establish the presence of the Quark Gluon
Plasma (QGP) and to determine its properties. Future physics
includes: 1) determining the extent of deconfinement; 2)
establishing a signature of chiral symmetry restoration; 3)
possibly distinguishing a strongly-coupled QGP from a
weakly-interacting one; 4) determining whether a color glass
condensate is formed in the initial state; and 5) understanding
parton propagation and the hadronization process.

Many exciting results have been presented at this Conference that
are consistent with formation of the QGP. However, in recent
overview and position papers on this topic, the experimental
collaborations at RHIC state that the evidence for discovery is
not presently conclusive
\cite{STAR_white,PHENIX_white,PHOBOS_white,BRAHMS_white}.
Theorists have tended to disagree with the experimentalists,
stating that the evidence exists \cite{Gyulassy_McLerran}. This
ongoing debate will be settled with data, and more detailed
calculations.

\section{Hard and EM Probe Physics at RHIC}

Jets, heavy flavors and electromagnetic probes will play a key
role in settling this debate, since they are sensitive to the
initial high density stage, when formation of a QGP is predicted.
Electromagnetic probes can be divided into two categories: direct
photons which tell us about thermal radiation and shadowing, and
virtual photons (electron-positron pairs) by way of their coupling
to vector mesons may tell us about chiral symmetry restoration and
possible bound states in a strongly-coupled QGP. Heavy flavors
include open charm, open beauty, and quarkonia (charmonium and
bottomonium states). Yields of quarkonia are predicted to be
strongly sensitive to deconfinement. Properties of jets can be
measured via leading particles, particle correlations, photon-jet
correlations, heavy quark (charm or beauty) tagged-jets, and
topological jet energy. These provide information on parton energy
loss, properties of the medium through which the partons
propagate, gluon shadowing, possible presence of a color glass
condensate, and hadronization mechanisms.

\subsection{RHIC Detector Capabilities}

PHENIX and STAR continue to improve triggering and data
acquisition capabilities in order to acquire data efficiently for
jets, heavy flavor and EM probes. New detector capabilities will
be added and apertures expanded. Additional detector capabilities
\cite{STAR_plan,PHENIX_plan} include micro-vertexing for
identifying displaced vertices from heavy flavor decays in STAR
($\mu$VTX) and PHENIX (MVTX), adding better low-mass
"hadron-blind" di-lepton capabilities in PHENIX (HBD), and
extending particle identification in STAR (ToF) and PHENIX
(Aerogel).

\subsection{RHIC Luminosities}

In order to undertake extensive studies of heavy flavor production
and jets at large transverse momenta (p$_T$), which have low cross
section (and are often called rare probes), an upgrade in the RHIC
luminosity will be necessary. The RHIC design luminosity for Au+Au
is L$_o$ = 2 $\times$ 10$^{26}$ cm$^{-2}$s$^{-1}$. RHIC now
routinely reaches twice this value. Thus, the anticipated $\int$
Ldt per RHIC year (20 weeks operation) is approximately 2 - 3
nb$^{-1}$. Note that the reference data for p+p (to understand
fundamental production mechanisms) and any comparison/control data
for d+Au (to understand normal nuclear effects) also require
statistics similar to those for Au+Au and extended RHIC running.

Many crucial measurements with hard and electromagnetic probes at
RHIC require $\int$ Ldt $>$ 20 nb$^{-1}$. For a vital program with
rare probes to continue at RHIC, the luminosity must increase. A
luminosity increase to 40 $\times$ L$_o$ or $\int$ Ldt $\sim$ 100
nb$^{-1}$ is planned for RHIC (and will be called RHIC II), with a
construction start possible in 2009 for operation in 2012.

\section{Electromagnetic Probes}

\subsection{Direct Photons}

Photons in A+A collisions may provide information on thermal
photon radiation. In p+A interactions, photons establish the
degree of shadowing. Photons in p+p reactions are needed for
reference data, to understand the underlying processes in p+A and
A+A results. Preliminary results on photons in Au+Au and p+p at
RHIC have been reported by PHENIX at this Conference
\cite{Reygers_HP04}. Photons measured in p+p are consistent with
next-to-leading order (NLO) pQCD calculations. Those measured in
Au+Au exhibit no thermal photons ($1 \leq p{_T} \leq 4$ GeV/c)
within present statistics. Furthermore, the Au+Au direct photons
are consistent with binary scaling of p+p. A definite statement
about direct photons in Au+Au at RHIC is anticipated from PHENIX
from the recent RHIC 2004 high statistics Au + Au run
\cite{Tserruya}.

\subsection{Virtual Photons via e$^+$e$^-$ pairs}

Thermal photons (measured via e$^+$e$^-$ pairs) are expected to be
radiated from the QGP.  However, at RHIC energies, the thermal
di-lepton spectrum in the intermediate mass range (1 - 3 GeV) may
be dominated by charm. In addition to information on thermal
radiation, virtual photons (measured via e$^+$e$^-$ pairs)
investigate possible modifications of vector mesons in the medium.
The behavior of vector mesons in medium may shed light on the
existence of chiral symmetry breaking and/or bound states in a
strongly-coupled QGP.

\begin{figure}
\centering
\includegraphics[height=7cm]{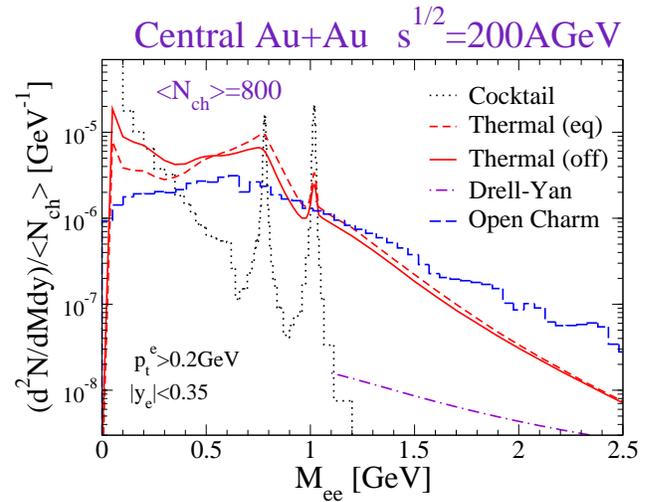}
\caption{Low mass e$^+$e$^-$ pair spectrum from \cite{Rapp}.
Contributions from the hadronic "cocktail", thermal and
non-equilibrium photons, Drell-Yan, and open charm are displayed
as denoted in the figure.} \label{fig:1}
\end{figure}

In order to effectively pursue low mass electron pair
measurements, PHENIX has proposed to install a hadron-blind TPC
(HBD) and STAR has proposed a barrel Time-of-Flight detector (ToF)
for electron identification at p$_T$ $>$ 0.2 GeV/c. A calculation
of the light vector meson yield as a function of invariant mass of
e$^+$e$^-$ pairs is displayed in Fig. \ref{fig:1} \cite{Rapp}.
Peaks for the $\rho$ and $\phi$ mesons are observed, but are
swamped by e$^+$e$^-$ pairs from thermal and non-equilibrium
photons and open charm. Careful investigation of medium
modification of low mass vector mesons requires measuring and
understanding all contributions to the di-lepton spectrum
including detailed charm studies that require RHIC II
luminosities.

\subsection{Heavy Flavor (Quarkonium)}

The production of quarkonium states in p+p, p+A, and A+A
collisions provides a tool to study deconfinement in strongly
interacting matter \cite{Matsui}. Studies of the dependence of the
heavy-quark potential on the in-medium temperature in lattice QCD
calculations with dynamical quarks \cite{Digal} indicate a
sequence of melting of the quarkonium states based upon their
binding strengths: T($\psi$') $<$ T($\Upsilon_{3S}$) $<$
T(J/$\psi$) $\sim$ T($\Upsilon_{2S}$) $<$ T($\Upsilon_{1S}$) where
T($\Upsilon_{3S}$) $<$ T$_c$ and T($\Upsilon_{1S}$) $>$ T$_c$,
with T$_c$ the deconfinement phase transition temperature.
Therefore, a measurement of the yields of the various bottomonium
states will shed light on the production (via $\Upsilon_{1S}$) and
suppression mechanisms ($\Upsilon_{2S}$ and $\Upsilon_{3S}$) of
quarkonia avoiding many difficulties inherent in charmonium
measurements. These measurements are challenging, requiring
excellent momentum resolution to resolve the bottomonium states
and very high rate (luminosity) and trigger capabilities because
of the low production cross-sections.

The larger production cross-sections for charmonium states
compared to bottomonium states have led to studies of charmonium
and the subsequent observation of charmonium suppression in
collisions of heavy ions at the SPS. PHENIX anticipates its first
results on charmonium suppression in Au+Au at RHIC from the large
statistics 2004 data run. Bottomonium spectroscopy on the other
hand requires higher luminosities. Since bottomonium is massive
($\sim$ 10 GeV/c$^2$) its decay leptons have sufficiently large
momenta above background processes facilitating high-level
triggering.

The PHENIX mass resolution for $\Upsilon$ $\rightarrow e^{+}e^{-}$
with the VTX detector upgrade is $\Delta$m = 60 MeV. Without the
VTX it is 170 MeV, making resolution of the $\Upsilon_{1S}$ (9.460
GeV), $\Upsilon_{2S}$ (10.020 GeV), and $\Upsilon_{3S}$ (10.360
GeV) challenging. The PHENIX mass resolution in the muon arms is
worse than 170 MeV making it difficult to resolve the individual
$\Upsilon$ states in the muon decay channel. Statistics for the
$\Upsilon$ states, combined in Table 1, are low. In STAR, the mass
resolution for $\Upsilon$ $\rightarrow e^{+}e^{-}$ is $\Delta$m =
340 MeV using the time projection chamber tracking alone. A
$\mu$-vertex detector upgrade would improve this resolution to
$\Delta$m = 170 MeV. Only with a planned data acquisition system
upgrade, will STAR be able to detect a significant number of
$\Upsilon$'s (1750 combined in all three states with 1.5 nb$^{-1}$
Au+Au). In general, a meaningful bottomonium program at RHIC will
require RHIC II luminosities ($\sim$ 100 nb$^{-1}$) and large
acceptances to obtain reasonable statistics.

The quarkonium statistics anticipated for Au + Au in PHENIX
\cite{PHENIX_plan,Drees} are presented in Table 1.

\begin{table}[h]
\centering \caption{PHENIX Quarkonium Program for Au+Au at RHIC
and RHIC II. \cite{PHENIX_plan,Drees}} \label{tab:1}
\begin{tabular}{lll}
\hline\noalign{\smallskip}
Channel   & RHIC (1.5 nb$^{-1}$)    & RHIC II (30 nb$^{-1}$) \\
\noalign{\smallskip}\hline\noalign{\smallskip}
J/$\psi \rightarrow$ e$^+$e$^-$ & 2,800 & 56,000 \\
$\psi$' $\rightarrow$ e$^+$e$^-$ & 100 & 2,000 \\
$\Upsilon$ $\rightarrow$ e$^{+}$e$^{-}$ & 8$\dag$ & 155$\dag$\\
(all states)& & \\
   &  &  \\
J/$\psi$ ($\psi$') $\rightarrow \mu^{+} \mu ^{-}$ & 38,000 (1400) & 760,000 (28,000)\\
$\Upsilon$ $\rightarrow \mu^{+} \mu ^{-}$  & 35$\ddag$ & 700$\ddag$\\
(all states)& & \\
\noalign{\smallskip}\hline \\
\end{tabular}
\end{table}
\vspace {-.7cm} \hspace{-.65cm}
$\dag$ requires MVTX upgrade\\
$\ddag$ requires $\mu$-trigger
system upgrade.\\

\subsection{Open Heavy Flavor}

Heavy flavor (charm and bottom) yields are sensitive to the
initial gluon density and are important components of
understanding J/$\psi$ and $\Upsilon$ production. Measuring the
energy loss of a heavy quark in the medium will indicate whether
heavy quarks suffer less energy loss than light quarks in the
medium as predicted by the "dead-cone effect" \cite{Dokshitzer}.

Initial measurements of charm cross sections and charm flow have
been made by identifying single electrons above background in STAR
\cite{STAR_electrons} and PHENIX \cite{PHENIX_electrons}. These
results indicate that low p$_T$ open charm exhibits elliptic flow
and are preliminary at the time of this Conference. Significant
measurements up to moderate p$_T \sim$ 5 - 6 GeV of open charm and
open beauty decays can be made with $\sim$ 3 nb$^{-1}$ and
upgraded detectors in STAR ($\mu$-vertex, ToF) and PHENIX (VTX).

\section{Jets}

High p$_T$ particles and jets can be used to probe the quark-gluon
plasma (QGP), study its properties and gain a better understanding
of high density QCD and hadronization. Measuring the modifications
of fragmentation functions (FF) of partons traversing the QGP in
A+A collisions relative to p+p and p+A collisions should
distinguish characteristics of the QGP compared to those of a
nuclear medium. The RHIC energy regime appears to be ideal for
these studies. Recent measurements in the forward direction at
RHIC indicate possible gluon shadowing in the initial state at low
x. Therefore, measurements over a specific part of phase space
(e.g. forward- or mid-rapidities) selects the x region of the
dominant process of interest. The higher energy regime of the
Large Hadron Collider (LHC) will provide increased particle yields
at high p$_T$  and at low x.

\begin{figure}[htb]
\vspace*{-.3cm}
    \begin{center}
    \includegraphics[width=0.50\textwidth]{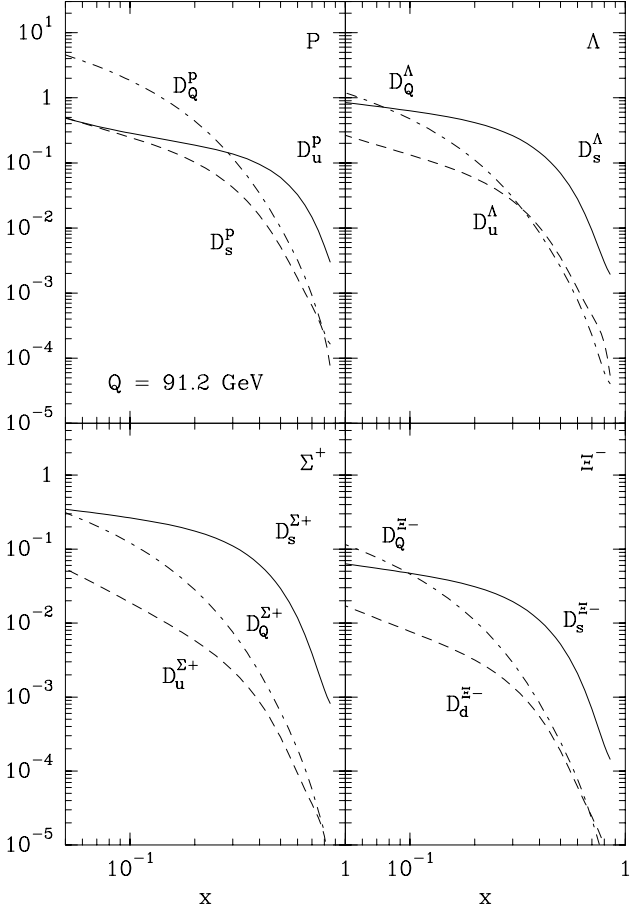}
    \vspace*{-0.3cm}
    \caption[]{Fragmentation functions (D$^{baryon}_{quark}$) for the octet
baryons (p, $\lambda$, $\Sigma$, and $\Xi$) as a function of
$\xBJ$ \cite{Bourrely-Soffer}. Contributions of the light (u,d),
strange (s), and heavy (Q = c, b) quarks are denoted by
subscripts.}
    \label{fig11}
    \end{center}
\end{figure}

The contributions of the various (u, d, s, c, b) quarks to the
mass of stable particles can be determined by measuring the
fragmentation function of each particle in $\pp$ interactions. The
contributions of the light (u,d,), strange (s), and heavy (Q =
c,b) quarks to the octet baryons (p, $\lambda$, $\Sigma$, and
$\Xi$) are presented as a function of $\xBJ$ in Fig. \ref{fig11}
\cite{Bourrely-Soffer}. Measurement of these fragmentation
functions requires particle identification of leading particles in
jets at large transverse momentum. Such measurements in $\NN$
collisions will establish how fragmentation functions are modified
by propagation of the various types of quarks in the dense medium
and will reflect these quark contributions to the particle masses
as they fragment in the medium. It would be extremely exciting if
fragmentation functions of some of the particles were to reflect
properties of a chirally restored medium. This latter connection
has yet to be established theoretically. In addition to accounting
for the constituent quark masses, the chiral quark condensate is
responsible for inducing transitions between left-handed and
right-handed quarks, $\qbar q$ = $\qbar_{L} q_{R}$ + $\qbar_{R}
q_{L}$. Therefore, helicities of (leading) particles in jets (e.g.
determined by detecting the polarization of leading $\Lambda$
particles) may provide information on parity violation and chiral
symmetry restoration \cite{Kharzeev-Sandweiss}.

Full utilization of hard parton scattering to probe high density
QCD matter requires measurements of photons (to establish the
parton energy), jets (another potential measurement of parton
energy), high p$_T$ identified particles (for fragmentation
functions of particles and flavor-tagging), and correlations from
among these. In addition, measurements are essential over a
multi-parameter space that can be divided into initial state (c.m.
energy, system mass, collision impact parameter, x$_1$ and x$_2$
of the colliding beam partons, and Q$^2$ of the collision) and
final state (p$_T^{parton}$, y$^{parton}$, $\phi^{parton}$,
p$_T^{jet/particle}$, y$^{jet/particle}$, $\phi^{jet/particle}$,
flavor$^{jet/particle}$, $\phi^{flow plane}$). Such a study
requires large data sets and high luminosity to extend
measurements to large p$_T$. Furthermore, it would be interesting
to investigate the difference in quark versus gluon propagation by
implementing kinematic cuts (x$_1$, x$_2$) in jet-jet
correlations, and to utilize the anticipated differences in the
yields of gluon and quark jets as a function of transverse
momentum and $\sqrt{s}$.

\subsection{Photon-tagged Jets}

The primary advantage of photons is that they do not re-interact
with the medium through which they propagate. Thus, they can be
used to determine the parton energy in the original
hard-scattering that produces the photon and away-side jet. The
away-side jet will suffer energy loss in the medium and thus the
difference of the photon and measured away-side jet energy can be
used to determine the energy lost by the parton on the away-side
of the photon. Correlation of the flavor of the leading hadron in
the jet on the away-side of a photon provides additional
information with which to test energy loss mechanisms. Another
advantage of photon-jet correlations is that there are only two
production diagrams that produce photons in leading order:
quark-antiquark annihilation and quark-gluon Compton scattering.
However, there is one major issue to be dealt with when utilizing
prompt photons to determine the momentum of the hard-scattered
parton. One must distinguish direct (prompt) photons from those
from fragmentation of partons. This is complicated and must be
resolved through understanding the contribution of fragmentation
photons and their momentum dependence.

STAR will undertake initial studies of photon-jet correlations up
to 10 GeV/c photon momentum by utilizing a few nb$^{-1}$ integral
luminosity in Au+Au at RHIC. Since less than 1 percent of the jets
have a leading hadron above the background in a Au+Au collision at
RHIC, a few year Au+Au run with 4 - 5 nb$^{-1}$ integral
luminosity is expected to yield $\sim$ 8K charged hadrons in a
spectrum on the away-side from a 10 GeV/c photon, and $\sim$ 1K
charged hadrons in a spectrum on the away-side from a 15 GeV/c
photon in STAR. More detailed measurements, especially with
identified particles on the away-side for fragmentation function
modification for high p$_T$ partons requires RHIC II luminosities.

PHENIX proposes to perform statistical photon-jet correlation
analyses with approximately 1000 photon-jet events. The maximum
photon p$_T$ depends on the PHENIX detector complement. In the
2004 Au+Au run the maximum photon p$_T$ (p$_T^{max}(\gamma)$) is
expected to be $\sim$ 6 GeV/c; with a new time projection chamber
(covering $-1 \leq \eta \leq 1$) at RHIC luminosity
p$_T^{max}(\gamma) \sim$ 12 GeV/c; with an additional nose cone
calorimeter for expanded photon detection at RHIC II luminosity
p$_T^{max}(\gamma) \sim$ 23 GeV/c at mid-rapidity \cite{Drees}.

As we have seen in this Conference \cite{Magestro}, jets in Au+Au
collisions broaden significantly in pseudorapidity on the
near-side as well as on the away-side. This can be attributed both
to the variation of parton momentum fractions for partons of the
incoming beam nuclei (parton momentum fractions x$_1$,x$_2$) and
to possible quenching in matter. Calculations utilizing PYTHIA 6.2
show that photons from $\sqrt s$ = 200 GeV p+p interactions at
RHIC extend with a fairly flat distribution over the
pseudo-rapidity range $-3 \leq \eta \leq 3$. The same simulations
show that the difference in  pseudo-rapidity between the photon
and away-side jet has a width $\sigma$($\eta_{\gamma} -
\eta_{jet}$) = 1.5 units of pseudo-rapidity. Experiments seeking
to undertake photon-jet measurements should have large acceptance
for photons, high p$_T$ particles and jet energy in order to cover
a range of parton momentum fractions (x$_1$,x$_2$).

\subsection{Flavor-tagged Jets}

Jets with a D- or B-meson as leading particles will serve to study
the response of the medium to heavy quarks compared to light ones.
A high p$_T$ electron in coincidence with a leading hadron, both
emanating from a vertex that is displaced from the primary
reaction vertex, will provide a trigger for heavy flavor decays.
Significant measurements at large p$_T \sim$ 10 - 15 GeV of open
charm and open beauty as leading particles of jets require RHIC II
luminosities and upgraded detectors in STAR ($\mu$-vertex, ToF)
and PHENIX (VTX).

%\begin{figure}[h]
%\vspace*{-.4cm}
%    \begin{center}
%    \includegraphics[width=0.5\textwidth]{fig2_exp_setup.eps}
%    \vspace*{-0.5cm}
%    \caption[]{Diagram of the new detector using the SLD magnet.}
%    \label{fig1}
%    \end{center}
%\end{figure}

\begin{figure*}[htb]
\centerline{\includegraphics[width=12cm]{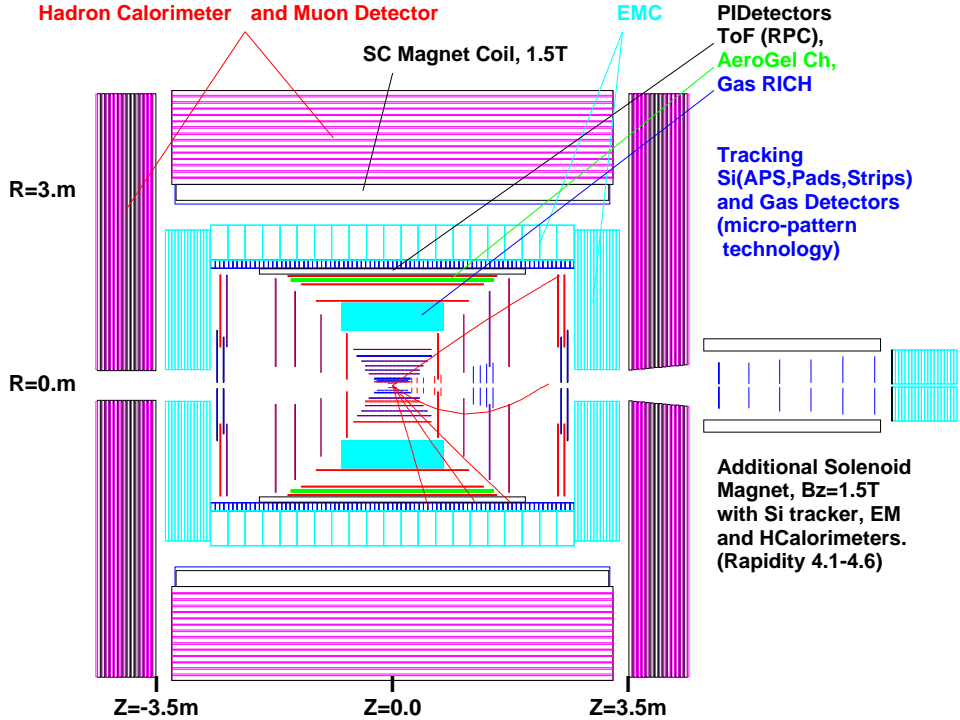}}
\caption{\small{Diagram of a possible, comprehensive new detector
at RHIC II using the SLD magnet.}} \label{fig1}
\end{figure*}

\section{A Comprehensive New RHIC II Detector}

A comprehensive new detector has been proposed
\cite{new_detector_EoI} for RHIC II to undertake measurements of
jets, heavy flavors, and electromagnetic probes, and to take full
advantage of the high RHIC II luminosities. New RHIC II physics
opportunities can be studied by utilizing a high field ($\sim$ 1.5
T) magnet, extensive charged hadron tracking and identification,
electron and muon tracking and identification, and extensive
coverage of electromagnetic and hadronic calorimetry. The
capabilities of a comprehensive new detector include: 1) excellent
charged particle momentum resolution to p$_T$ = 40 GeV/c in the
central rapidity region, 2) complete hadronic and electromagnetic
calorimetry over a large phase space (-3 $\leq$ $\eta$ $\leq$ 3,
$\Delta \phi$ = 2$\pi$), 3) particle identification out to large
p$_T$ ($\sim$ 20 - 30 GeV/c) including hadron ($\pi$, K, p) and
lepton (e/h, $\pi$/h) separation in the central and forward
region, and 4) high rate detectors, data acquisition, and trigger
capabilities. A possible layout for a new RHIC II detector using
the SLD magnet is shown in Fig. \ref{fig1}.

In order to identify all charged hadrons in a high $\pT$ jet, good
hadron identification is necessary up to momenta of approximately
20 GeV/c. Lepton particle identification will be achieved through
the e/h capabilities in the calorimeters and the muon chambers.
Hadron and lepton particle identification will be achieved through
a combination of dE/dx in the tracking detectors ($p_T$ $\leq$ 1
GeV/c), a time-of-flight device ($p_T$  $\leq$ 3 GeV/c), and a
combination of two different Aerogel Cherenkov-threshold counters
and a RICH detector with gas radiator  (up to $p_T$ $\sim$ 20
GeV/c). For more details on the comprehensive new detector see
\cite{new_detector_EoI}.

\subsection{Jets in a Comprehensive New Detector}

Jet rates over the large acceptance of a new detector (-3 $\leq$
$\eta$ $\leq$ 3, $\Delta \phi$ = 2$\pi$ plus extended forward
coverage to $\eta \sim$ 4.5) at upgraded RHIC II luminosities will
be significant. The anticipated jet yield for 40 GeV jets in such
a new detector at RHIC II with 30 nb$^{-1}$ of Au+Au at top energy
is $\sim$ 180,000. 19,000 $\gamma$-jet events are expected with
p$_{T}(\gamma)$ = 20 GeV/c, and 1,000 $\gamma$-jet events for
p$_{T}(\gamma)$ = 30 GeV/c with full away-side particle
identification over -3 $\leq$ $\eta$ $\leq$ 3 for determination of
the modification of fragmentation functions of particles.

Extension of high resolution particle tracking, particle
identification (PID), and calorimetry to forward rapidities will
be important in elucidating the various particle production and
hadronization mechanisms, which may be sensitive to the quark and
gluon components of the hadronic wave functions. At very low
$\xBJ$, gluons can be coherent over nuclear distances and a color
glass condensate \cite{McLerran,Kharzeev-Levin} formed. This would
have effects on many hard physics observables that depend directly
on the gluon structure, e.g. minijet rates and heavy flavor
production, but can be clarified by comparisons of $\pA$ physics
with $\pp$. Thus, it is important to study high $\pT$ processes
away from midrapidity. To take full advantage of physics in the
forward region, momentum measurements and PID must be undertaken
up to $\pT$ $\sim$ 2-3 GeV/c, which is a real experimental
challenge with longitudinal momenta of 20-30 GeV/c at large
rapidities.

\subsection{Quarkonia in a Comprehensive New Detector}

For determination of the quarkonium melting sequence an energy
resolution of better than 10$\%$/$\sqrt(E)$ is required to resolve
the quarkonium states with calorimeter information alone. Thus,
quarkonium physics at RHIC II in this new detector can be fully
realized by using the EMC in combination with high resolution
tracking and large acceptance muon chambers. The mass resolution
for $\Upsilon$ $\rightarrow \mu^{+} \mu^{-}$ in the new
comprehensive detector is $\Delta$m = 60 MeV. Furthermore, large
acceptance in the Feynman x$_F$ variable is important for
understanding quarkonium production and melting mechanisms. This
leads to the need for large acceptance in $\eta$ for lepton pairs.
The electron and muon coverage of the new detector extends over -3
$\leq$ $\eta$ $\leq$ 3 and $\Delta \phi$ = 2$\pi$. A similar
acceptance in the new detector for charmonium feed-down photons
from $\chi_c$ decays ($\chi_{c} \rightarrow$ J/$\psi$ + $\gamma$)
allows determination of the $\chi_c$ feed-down contribution to
J/$\psi$ production and subsequent suppression.

The anticipated quarkonium statistics for Au + Au in the new
comprehensive detector are presented in Table 2. These numbers for
the new detector are to be compared to those for PHENIX in Table
1.

\begin{table}
\centering \caption{New RHIC II Quarkonium Program for Au+Au at
RHIC II for p$_{lepton}$ $>$ 2 GeV/c for J/$\psi$, and
p$_{lepton}$ $>$ 4 GeV/c for $\Upsilon$} \label{tab:1}
\begin{tabular}{lll}
\hline\noalign{\smallskip}
Channel   & RHIC II (30 nb$^{-1}$) \\
\noalign{\smallskip}\hline\noalign{\smallskip}
J/$\psi \rightarrow$ di-leptons & 36,000,000 \\
$\psi$' $\rightarrow$ di-leptons & 1,000,000 \\
$\chi_c$' $\rightarrow$ J/$\psi$ + $\gamma$ & 680,000 \\
$\Upsilon$ $\rightarrow$ di-leptons & 64,000\\
$\Upsilon$' $\rightarrow$ di-leptons  & 12,000\\
$\Upsilon$'' $\rightarrow$ di-leptons  & 12,000\\
\noalign{\smallskip}\hline \\
\end{tabular}
\end{table}

\section{Conclusions}

The future of hard and electromagnetic probes at RHIC is "alive
and well"!  There is much new data still to be accumulated at
RHIC. From this data new physics will be uncovered, since jets,
heavy flavors and electromagnetic probes are sensitive to the
initial high density stage of RHIC collisions, when quark-gluon
plasma (QGP) formation is expected. Questions that still remain to
be addressed are whether 1) the system becomes deconfined, 2)
chiral symmetry is restored, 3) in addition to a strongly-coupled
QGP there is a weakly-interacting one, 4) a color glass condensate
is formed in the initial state, and 5) whether we can gain new
understanding of the hadronization process. Precise timescales for
new detector implementation to improve capabilities for rare
probes at RHIC is uncertain due to ambiguities in the availability
of funding. Significant capabilities will be added with new
detectors at RHIC and a possible comprehensive new detector at
RHIC II. We look forward to significant progress in these
directions and exciting physics from the hard and EM probe sector
at RHIC and RHIC II.

\section{Acknowledgements}

The author wishes to thank R. Bellwied, T. Ullrich, N. Smirnov, P.
Steinberg, H. Caines, M. Lamont, C. Markert, J. Sandweiss, M. Lisa
and D. Magestro for fruitful RHIC II physics discussions,
collaboration on the comprehensive new detector project, and
contributions to this work. M. Gyulassy, B. Mueller and D.
Kharzeev have contributed through enlightening discussions.

%

%%%%%%%%%%%%%%%%%%%%%%%%%%%%%%%%%%%%%%%%%%%%%%%%%%%%%%%%%%%%%%%%%%%%%%  }

%%%%%%%%%%%%%%%%%%%%%%%%%%%%%%%%%%%%%%%%%%%%%%%%%%%%%%%%%%%%%%%%%%%%%%

\end{document}